\documentclass[
 reprint,
 amsmath,amssymb,
 aps,
]{revtex4-1}

\usepackage{verbatim}
\usepackage{url}
\usepackage{graphicx}
\usepackage{xcolor}
\definecolor{lcblue}{HTML}{5CB8E8}
\newcommand{\dcz}[2]{\mathrm{CZ}_{#1\text{--}#2}}
\newcommand{\bpair}[2]{\mathrm{BP}_{#1\text{--}#2}}

\usepackage{dcolumn}
\usepackage{bm}

\renewcommand{\thetable}{\arabic{table}}

\makeatletter
\newcommand{\edlabels}{%
  \setcounter{figure}{0}\setcounter{table}{0}%
  \renewcommand{\figurename}{Extended Data Fig.}%
  \renewcommand{\tablename}{Extended Data Table}%
  \renewcommand{\thetable}{\arabic{table}}%
  \renewcommand{\thefigure}{\arabic{figure}}%
}
\makeatother

\begin{document}


\title{Meter-scale quantum interconnects between large quantum processors}

\author{Neereja M. Sundaresan}
\email{neereja@ibm.com}
\affiliation{IBM Research, T.~J. Watson Research Center, Yorktown Heights, NY 10598, USA}
\author{Nicholas Cotaj}
\affiliation{IBM Research, T.~J. Watson Research Center, Yorktown Heights, NY 10598, USA}
\author{Yadav Prasad Kandel}
\affiliation{IBM Research, T.~J. Watson Research Center, Yorktown Heights, NY 10598, USA}
\author{Eric Lewandowski}
\affiliation{IBM Research, T.~J. Watson Research Center, Yorktown Heights, NY 10598, USA}
\author{Christian Robert Longton}
\affiliation{IBM Research, T.~J. Watson Research Center, Yorktown Heights, NY 10598, USA}
\author{Yves Martin}
\affiliation{IBM Research, T.~J. Watson Research Center, Yorktown Heights, NY 10598, USA}
\author{Jae-woong Nah}
\affiliation{IBM Research, T.~J. Watson Research Center, Yorktown Heights, NY 10598, USA}
\author{Kathryn Pooley}
\affiliation{IBM Research, T.~J. Watson Research Center, Yorktown Heights, NY 10598, USA}
\author{Martin Sandberg}
\altaffiliation[Present address: ]{Anderon}
\affiliation{IBM Research, T.~J. Watson Research Center, Yorktown Heights, NY 10598, USA}
\author{Michael Scanlon}
\affiliation{IBM Research, T.~J. Watson Research Center, Yorktown Heights, NY 10598, USA}
\author{Chi Xiong}
\affiliation{IBM Research, T.~J. Watson Research Center, Yorktown Heights, NY 10598, USA}
\author{Emanuil Sashev Yanev}
\affiliation{IBM Research, T.~J. Watson Research Center, Yorktown Heights, NY 10598, USA}
\author{David Abraham}
\affiliation{IBM Research, T.~J. Watson Research Center, Yorktown Heights, NY 10598, USA}
\author{Abhinandan Antony}
\affiliation{IBM Research, T.~J. Watson Research Center, Yorktown Heights, NY 10598, USA}
\author{Vincent Arena}
\affiliation{IBM Research, T.~J. Watson Research Center, Yorktown Heights, NY 10598, USA}
\author{Nicholas Bornman}
\affiliation{IBM Research, T.~J. Watson Research Center, Yorktown Heights, NY 10598, USA}
\author{Khanh Dang}
\affiliation{IBM Research, T.~J. Watson Research Center, Yorktown Heights, NY 10598, USA}
\author{Oliver Dial}
\affiliation{IBM Research, T.~J. Watson Research Center, Yorktown Heights, NY 10598, USA}
\author{Christopher Guarino}
\affiliation{IBM Research, T.~J. Watson Research Center, Yorktown Heights, NY 10598, USA}
\author{Gabriel Haspil}
\affiliation{IBM Research, T.~J. Watson Research Center, Yorktown Heights, NY 10598, USA}
\author{Jared Hertzberg}
\affiliation{IBM Research, T.~J. Watson Research Center, Yorktown Heights, NY 10598, USA}
\author{Kentaro Heya}
\affiliation{IBM Research, T.~J. Watson Research Center, Yorktown Heights, NY 10598, USA}
\author{Swetha Kamlapurkar}
\affiliation{IBM Research, T.~J. Watson Research Center, Yorktown Heights, NY 10598, USA}
\author{Michael Karschner}
\affiliation{IBM Research, T.~J. Watson Research Center, Yorktown Heights, NY 10598, USA}
\author{George Keefe}
\affiliation{IBM Research, T.~J. Watson Research Center, Yorktown Heights, NY 10598, USA}
\author{Eric Liniger}
\affiliation{IBM Research, T.~J. Watson Research Center, Yorktown Heights, NY 10598, USA}
\author{Laura Luo}
\affiliation{IBM Research, T.~J. Watson Research Center, Yorktown Heights, NY 10598, USA}
\author{Thomas G. McConkey}
\altaffiliation[Present address: ]{Anderon}
\affiliation{IBM Research, T.~J. Watson Research Center, Yorktown Heights, NY 10598, USA}
\author{David Olson}
\affiliation{IBM Research, T.~J. Watson Research Center, Yorktown Heights, NY 10598, USA}
\author{Jason Orcutt}
\altaffiliation[Present address: ]{Anderon}
\affiliation{IBM Research, T.~J. Watson Research Center, Yorktown Heights, NY 10598, USA}
\author{Dmitry Ovchinnikov}
\affiliation{IBM Research, T.~J. Watson Research Center, Yorktown Heights, NY 10598, USA}
\author{Timothy Phung}
\altaffiliation[Present address: ]{Amazon Web Services}
\affiliation{IBM Research, T.~J. Watson Research Center, Yorktown Heights, NY 10598, USA}
\author{Sankeerth Rajalingam}
\affiliation{IBM Research, T.~J. Watson Research Center, Yorktown Heights, NY 10598, USA}
\author{Michael Selvanayagam}
\affiliation{IBM Research, T.~J. Watson Research Center, Yorktown Heights, NY 10598, USA}
\author{William Shanks}
\affiliation{IBM Research, T.~J. Watson Research Center, Yorktown Heights, NY 10598, USA}
\author{Dongbing Shao}
\affiliation{IBM Research, T.~J. Watson Research Center, Yorktown Heights, NY 10598, USA}
\author{Abigail Shearrow}
\affiliation{IBM Research, T.~J. Watson Research Center, Yorktown Heights, NY 10598, USA}
\author{Adam Stabile}
\affiliation{IBM Research, T.~J. Watson Research Center, Yorktown Heights, NY 10598, USA}
\author{Jiri Stehlik}
\affiliation{IBM Research, T.~J. Watson Research Center, Yorktown Heights, NY 10598, USA}
\author{Joseph Suttle}
\affiliation{IBM Research, T.~J. Watson Research Center, Yorktown Heights, NY 10598, USA}
\author{Jeng-Bang Yau}
\affiliation{IBM Research, T.~J. Watson Research Center, Yorktown Heights, NY 10598, USA}
\author{Eric Zhang}
\affiliation{IBM Research, T.~J. Watson Research Center, Yorktown Heights, NY 10598, USA}

\date{\today}

\begin{abstract} Modular quantum architecture requires long-distance quantum interconnects to stitch together
individual processor modules. Such interconnects, long-distance couplers or L-couplers, must support coherent
quantum operation over meter-scale distances~\cite{zhong2021deterministic,magnard2020link,storz2023bell} and
scale to multiple connections between any two modules~\cite{yoder2025tourdegross}. Here
we demonstrate four L-couplers operating between two 120-qubit IBM Nighthawk R1 superconducting qubit
processors. Each L-coupler joins two Nighthawk lattice qubits through two frequency-tunable L-qubits, two
RF-SQUID based ``i2m'' couplers, and 80\,cm of superconducting aluminum coax~\cite{martin2024indium}. Each cable installs
perpendicular to the processor surface, to enable denser connections, and is joined to the chip by an
indium-to-indium room temperature pressure bond made after both modules were mounted.
L-coupler modes near 5\,GHz, with internal quality factors of $0.86$ to
$1.09\times10^{6}$, carry relay-based state transfer~\cite{zhong2021deterministic} at a mean
network benchmarking~\cite{helsen2023benchmarking} fidelity of $99.00\%$, with a best link of
$99.29 \pm 0.07\%$. We then realize distant CZ gates~\cite{heya2025rb} between lattice qubits in separate
modules at a mean fidelity of $94.75\%$, with a best link of $96.42 \pm 0.28\%$, and use them
to generate entangled states across the
two processors, including a demonstration of simultaneous Bell pair generation using two
adjacent L-couplers. These four
high-fidelity L-couplers establish a practical path to scalable quantum interconnects for
modular quantum computing.
\end{abstract}

\maketitle


\begin{figure*}[t]
    \centering
    \includegraphics[width=\linewidth]{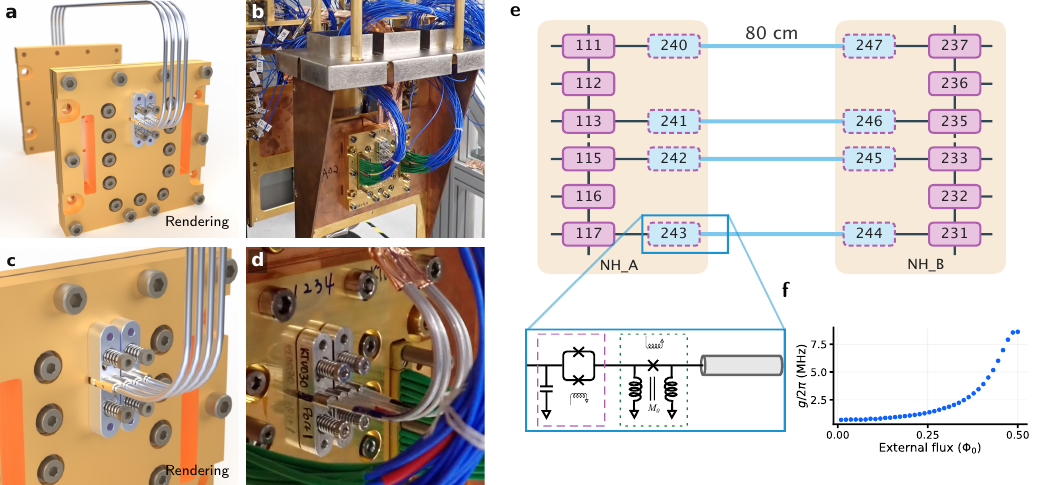}

    \caption{Two Nighthawk R1 test modules connected by four L-couplers.
a,b) Rendering and photograph of the dual-module assembly.
c,d) Rendering and photograph of the spring-loaded cable holder, which installs two L-coupler
connections at a time.
e) For development speed, only a subset of qubits around the L-couplers is wired and
measured. Solid outlined boxes are lattice qubits; the dashed box is one L-coupler element,
which the inset expands: a dedicated grounded, frequency-tunable L-qubit, an i2m coupler, and
the connection to the cable.
f) Representative measurement of qubit-cable coupling for one L-coupler versus external
flux.
See Methods for details.}
    \label{fig:Fig1}
\end{figure*}

A modular quantum computer comprises individual quantum processor modules joined by
long-distance quantum interconnects. In our superconducting qubit architecture, long-distance
couplers called L-couplers form the cryogenic quantum interconnects, operating at microwave
frequencies. Within each module a dense lattice gives high-degree connectivity and the
highest available two-qubit gate fidelity, while between modules a sparser set of L-coupled
edges carries the entanglement required for the collection to operate as a single machine.
Partitioning a machine this way helps even when those sparse inter-module links carry error
rates an order of magnitude larger than gates within the
lattice~\cite{yoder2025tourdegross,nickerson2013noisy,ramette2024interface,marqversen2025interfaces},
and can reduce the physical qubit overhead of error correction by an order of
magnitude~\cite{higgott2026modular}. The relevant length scale for the
L-coupler is of order a meter, set predominantly by the module cryogenic environment and by classical
signal delivery~\cite{ibm2026cryo,schar2026link}.

To enable high fidelity intermodule operation, the L-coupler must be low loss and high
coherence. Superconducting aluminum coaxial cables have been shown to reach internal quality
factor $Q_i \sim 1.5$--$2\times10^{6}$ at 5 GHz~\cite{martin2024indium}. A major obstacle is
connecting the cable to the processor chip in a way that is compatible with how a large, fully
wired processor is built and installed. L-coupler counts are expected to increase substantially
over the coming years, which will stress reproducibility, scalability and yield.

Module-to-module quantum operation over superconducting cables has been demonstrated in many forms, including
meter-scale aluminum coaxial links between processors, multi-module state transfer and
entanglement, teleportation over cable buses tens of meters long,
links between separate cryogenic systems, and quantum state routers among detachable
modules~\cite{zhong2019bell,zhong2021deterministic,niu2023lowloss,song2024entangler,qiu2025teleportation,magnard2020link,storz2023bell,zhou2023router}.
Many of these demonstrations used few-qubit modules, with a single L-coupler between any
pair of processors, and relied on bench-top tooling to attach the cable to the chip.
A detachable cable connection between superconducting qubit modules was separately
demonstrated, reaching inter-module SWAP at the $99\%$ level~\cite{mollenhauer2025plugplay},
but in an architecture not readily compatible with the large-qubit processor used here.
In another scheme, deterministic remote entanglement was shown over a chiral quantum
link~\cite{almanakly2025chiral}. While remote two-qubit gates over a cable have been proposed
in several forms~\cite{ohfuchi2024crossres,yuan2026nonlocalzz,xu2026remotecz}, the relay-based
state transfer of Zhong \textit{et al.}~\cite{zhong2021deterministic} is one of the most direct
routes to operating a cable as a quantum link, and is what we adopt here.

Our hardware approach builds on the indium-to-indium room temperature pressure bond of our earlier
work~\cite{martin2024indium}, which made a high quality cable-to-chip connection using
bench-top techniques, and a cable-to-cable bond that could be assembled in place. In our
Flamingo builds~\cite{ibm2024qdc}, 100+ qubit modules with four L-couplers, we built on
this for a cable-to-chip join parallel to the processor surface, plugging in one cable at a
time. That work established that indium joins assembled in place are viable. However, the
reliability and reproducibility of those connections were limiting, and dense scaling to
larger cable count an issue. To address those key issues, we arrived at
the perpendicular cable-to-chip join geometry (Fig.~\ref{fig:Fig1}a--d).

Here we report four L-couplers between two 120-qubit processors, assembled at room
temperature after both modules are mounted in the cryostat. They reach a mean $Q_i$ of the
L-coupler modes near $5$~GHz of $0.94\times10^{6}$ and a mean state transfer fidelity of
$99.00\%$. Using state transfer, we construct distant CZ gates between lattice qubits on
separate processors at a mean fidelity of $94.75\%$, and use them to distribute Bell and GHZ
states across the two modules, including a demonstration of simultaneous Bell pair generation
on two neighbouring L-couplers.

\section*{The L-coupler}

We connect two IBM Nighthawk R1~\cite{ibm2025nighthawk} superconducting
qubit processors with four L-couplers, each with a continuous $\sim 80$\,cm superconducting
aluminum coaxial cable between modules (Fig.~\ref{fig:Fig1}a,b).

Our implementation uses two types of qubits: fixed-frequency Nighthawk lattice qubits, and
dedicated frequency-tunable L-qubits (Fig.~\ref{fig:Fig1}e). We define the L-coupler to include the dedicated
quantum-chip elements: the L-qubits, the tunable qubit-cable couplers, the on-chip trace and
transitions, the physical attachment of the coax to the chip, and the coax itself. Each
L-coupler is connected to a lattice qubit through a tunable coupler that enables a local CZ
gate. At each end the L-qubit is galvanically connected to the tunable qubit-cable coupler,
which couples it to the cable (Fig.~\ref{fig:Fig1}e,f). We call this coupler the i2m; it is descended from the
RF-SQUID based design~\cite{chen2014gmon}, and its always-on negative geometric mutual
inductance makes the off-coupling point first-order flux insensitive (Methods).

\section*{Magnetic sensitivity of aluminum cables}

\begin{figure}[t]
    \centering
    \includegraphics[width=\linewidth]{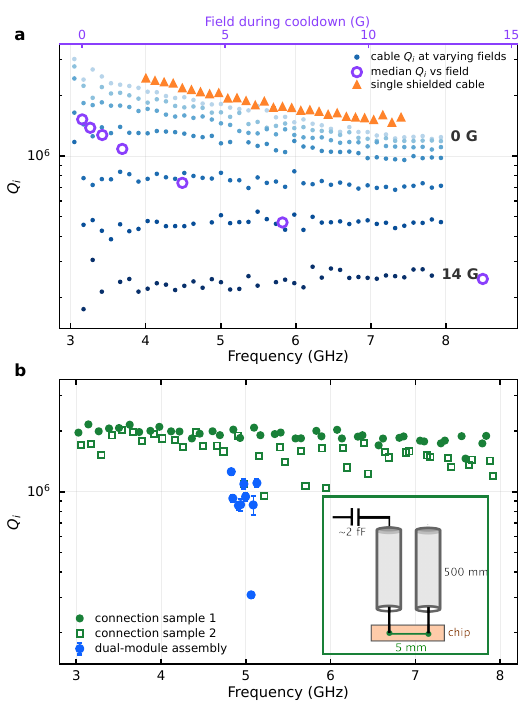}
    \caption{Magnetic sensitivity of aluminum cables and characterization of cable-chip
connections.
a) $Q_i$ of a standalone aluminum coax against frequency, for a series of fields applied
while the cable cooled through its superconducting transition, shaded light to dark from zero
to $14$~G and labelled at the extremes. Purple open circles give the median $Q_i$ across the
$4.5$--$7.5$~GHz band at each field, read against the upper axis. Orange triangles are a
single shielded cable on the route used in the dual-module assembly.
b) Green circles and open squares are two nominally identical standalone cable-to-chip join
testers. Blue circles are the four cable modes of the dual-module assembly, inferred from the
L-qubits. Inset: the standalone tester, in which two cable-to-chip joins sit near the midpoint
of a resonator formed from $500$~mm of coax either side of a short on-chip trace.
Error bars are fit uncertainties. See Methods for details.}
    \label{fig:Fig2}
\end{figure}

Module magnetic shields protect each processor, but the L-coupler cable necessarily extends
outside them, routing through regions of elevated magnetic field. We
show data that identified the magnetic field present while the cable cools through its
superconducting transition as a significant and previously undocumented loss channel for this class of
interconnect (Fig.~\ref{fig:Fig2}a). Taking the median $Q_i$ ($4.5$--$7.5$~GHz) at each cooldown field,
we find $1.5\times10^{6}$ at zero applied field, $1.1\times10^{6}$ at $1.4$~G,
$7.4\times10^{5}$ at $3.5$~G and $2.5\times10^{5}$ at $14$~G, a factor of six of loss across
the range probed. Changing the field after the cable had
cooled below $T_c$ had no effect, so the loss is set during the transit through the transition
and cannot be recovered by compensating the field at base temperature. Fields of this
magnitude are neither unusual nor well controlled between modules, so an unshielded cable can
arrive at base temperature already degraded.

Magnetic shielding of the coax is therefore essential. We mitigate the effect with a
shielding braid around the cable~\cite{magshield2026aacable}, which we regard as a temporary
solution: a single layer did not suffice, so we stacked two layers (Methods). To test it we cooled one
shielded cable along the route the assembly uses, where the field is uncontrolled. It retains
a median $Q_i$ of $2.0\times10^{6}$ near $5$~GHz (Fig.~\ref{fig:Fig2}a, orange), as high as a
bare cable cooled in zero applied field. In the assembled pair all four cables share a single
braid, a related but distinct shielding configuration from the single-cable test, and there
the four cable modes reach $Q_i = 0.95$, $1.09$, $0.86$ and $0.86
\times 10^{6}$ near $5$~GHz, inferred from the L-qubits, a mean of $(0.94 \pm
0.09)\times10^{6}$ with the spread taken across links. Even bundled this way they exceed the
highest values published for an assembled
interconnect~\cite{niu2023lowloss,mollenhauer2025plugplay}, and the mode dephasing is near
the decay limit, indicating good frequency stability.

\section*{Cable-to-chip connection}

In this demonstration we use a custom spring-loaded cable holder containing two cables and
install two connections at once (Fig.~\ref{fig:Fig1}c,d), made after both modules were mounted
in the cryostat using the holder's screws and visual alignment alone.
We install two cables in a single row
here, but the geometry is readily extendable to 2D packing.

Our ability to study wide-band properties of the cable through the L-qubit is limited, so we
developed and proved out the cable connection in a standalone tester, which places two
cable-to-chip joins near the midpoint of a resonator formed from $500$~mm of coax on either
side of a short on-chip trace (Fig.~\ref{fig:Fig2}b), a comprehensive way to study cable-chip
joins~\cite{martin2024indium}. Across two test samples we measure a mean $Q_i$ of $1.74\times10^{6}$ over
$3$--$8$~GHz, on par with the $1.5$--$2\times10^{6}$ of bare aluminum coax across the full
band: the join itself adds little measurable loss over that range. All four L-couplers
remained fully operational across three thermal cycles with all connections left untouched.
A modular machine has to be cooled, warmed and reworked many times over its life, so
connections that survive thermal cycling without intervention are a practical prerequisite for
modular quantum computing.

\section*{State transfer, distant gates, and entanglement}

\begin{figure}[t]
    \centering
    \includegraphics[width=\linewidth]{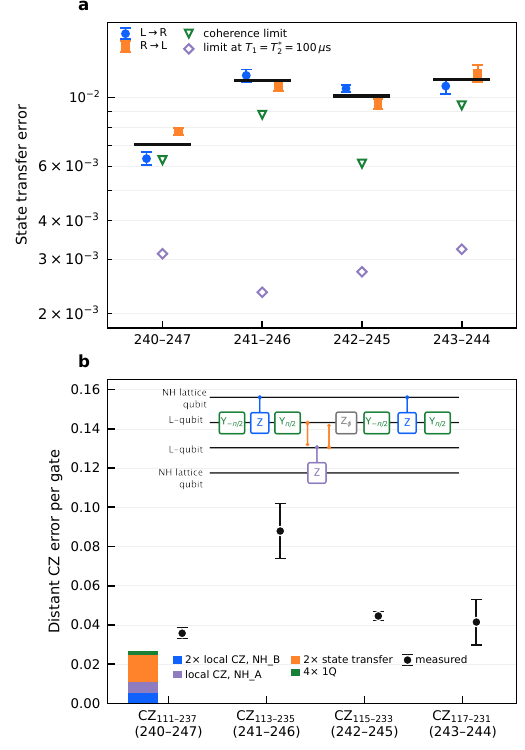}
    \caption{State transfer and distant CZ on L-couplers.
a) Relay-based state transfer error per L-coupler, measured in both directions with network
benchmarking, with the mean as a bar. Open triangles are the coherence limit of the model in
Methods, which takes the shorter measured $T_1$ and the shorter $T_2^{*}$ of each L-qubit pair
together with the measured L-coupler mode $T_1$ and $T_2^{*}$; open diamonds are the same model
with the L-qubits held at $T_1 = T_2^{*} = 100$\,$\mu$s.
b) Distant CZ error per gate, labelled by the outer lattice qubits the gate acts on with the
mediating L-coupler in parentheses. The stacked bar is the summed constituent error budget for
$\dcz{111}{237}$, beside its measured point. Inset: the gate circuit, in which orange arrows
mark the two state transfers.
Error bars are fit or shot-noise uncertainties. Per-link numbers are in Extended Data
Tables 1 and 2.}
    \label{fig:Fig3}
\end{figure}

At 80\,cm the L-coupler free spectral range is about $150$~MHz, and the modules here realize a
modest $g/\mathrm{FSR} = 0.05$--$0.06$. A single standing mode of the coax near $5$~GHz can
therefore be isolated, and we use it for relay-based state transfer between the two L-qubits of
an L-coupler~\cite{zhong2021deterministic}. Transfer folds L-qubit coherence, qubit-cable
coupling and cable loss into one number, which we measure by network
benchmarking~\cite{helsen2023benchmarking,heya2025rb}. All four L-couplers transfer at or above
$98.86\%$, with a mean of $99.00\%$ and a best link of $99.29 \pm 0.07\%$
(Fig.~\ref{fig:Fig3}a).

A simple error model including only L-qubit and cable mode coherences, and so an upper bound
on fidelity, accounts for $77\%$ of the measured transfer error, with L-qubit dephasing
dominating that. Holding the L-qubits at $T_1 = T_2^{*} = 100$\,$\mu$s the same model bounds
fidelity between $99.68\%$ and $99.77\%$ on all four links (Methods). As we intentionally skipped fabrication and processing steps known to
improve qubit coherence for development speed, the limited qubit coherence here is not a
reflection of L-coupling and can therefore be readily improved.

We construct a distant CZ gate between two Nighthawk lattice qubits in different modules,
mediated by an L-coupler. Following Heya \textit{et al.}~\cite{heya2025rb}, it combines three local CZ
gates, two state transfers, four single-qubit gates and one virtual phase correction
(Fig.~\ref{fig:Fig3}b). We
benchmark distant CZ gates on every L-coupler of the assembly (Fig.~\ref{fig:Fig3}b). Three
reach $95.54 \pm 0.23\%$ or better and the best is $96.42 \pm 0.28\%$ on $\dcz{111}{237}$
through the $240$--$247$ L-coupler, which is also the best transfer link. A simple budget of
the measured constituents upper-bounds that gate at $97.34 \pm 0.05\%$, dominated by the two
state transfers; we attribute much of the remaining discrepancy to residual population left
on the L-qubits by incomplete state transfer (Methods).

With the distant CZ we generate entangled states on lattice qubits spanning the two modules.
Created individually, the Bell pair $\bpair{115}{233}$ through the $242$--$245$ L-coupler reaches
$F = 85.91 \pm 0.40\%$ ($93.94 \pm 0.27\%$ mitigated) and $\bpair{117}{231}$ through
$243$--$244$ reaches $76.93 \pm 0.49\%$ ($86.98 \pm 0.41\%$). We can generate these Bell pairs
simultaneously using both L-couplers, a proof that the L-couplers here can be operated
independently (Table~\ref{tab:ent}). Finally we extend beyond the Bell pair to generate
four- and six-qubit GHZ states on lattice qubits, at $F = 77.23 \pm 0.67\%$ ($91.70 \pm
0.70\%$) and $67.67 \pm 0.66\%$ ($90.87 \pm 0.74\%$), demonstrating L-coupler enabled
entangling operations beyond the immediate neighbourhood of the L-qubits in the Nighthawk
modules, an important milestone for large-scale qLDPC code implementation.

\begin{table}[htbp]
\centering
\caption{Entangled states distributed across the two modules, as state fidelities with and
without readout mitigation. $F > 50\%$ certifies entanglement for the Bell
pairs and genuine multipartite entanglement for the GHZ states. Each Bell pair is given isolated,
with the adjacent L-coupler idle, and with both pairs generated simultaneously.
Figure~\ref{fig:Fig1}e gives the topology.}
\label{tab:ent}
\begin{ruledtabular}
\begin{tabular}{lcc}
 & \multicolumn{2}{c}{$F$ (\%)} \\
\cline{2-3}
State & unmitigated & mitigated \\
\colrule
\multicolumn{3}{l}{$\bpair{115}{233}$ via $242$--$245$} \\
\quad isolated      & $85.91 \pm 0.40$ & $93.94 \pm 0.27$ \\
\quad simultaneous  & $80.40 \pm 0.46$ & $87.56 \pm 0.38$ \\[2pt]
\multicolumn{3}{l}{$\bpair{117}{231}$ via $243$--$244$} \\
\quad isolated      & $76.93 \pm 0.49$ & $86.98 \pm 0.41$ \\
\quad simultaneous  & $76.17 \pm 0.49$ & $86.14 \pm 0.58$ \\[2pt]
\multicolumn{3}{l}{GHZ via $242$--$245$} \\
\quad $Q115$, $Q116$, $Q232$, $Q233$ & $77.23 \pm 0.67$ & $91.70 \pm 0.70$ \\
\quad $Q115$--$Q117$, $Q231$--$Q233$ & $67.67 \pm 0.66$ & $90.87 \pm 0.74$ \\
\end{tabular}
\end{ruledtabular}
\end{table}

\section*{Discussion}

In the coming years quantum computers will scale up markedly in both processor complexity and
module count. Such machines will need many nominally identical connections between
modules~\cite{yoder2025tourdegross,higgott2026modular}.

We demonstrate a practical route to realizing multiple L-couplers between large-qubit
processors after the modules are mounted in a cryostat. Across the four L-couplers reported here
the mean cable mode $Q_i$ is $0.94\times10^{6}$, the mean state transfer fidelity $99.00\%$ and
the mean distant CZ fidelity $94.75\%$, between lattice qubits in separate processors. To our
knowledge this is the first demonstration of a multi-channel quantum interconnect between
large-qubit superconducting processors, and it establishes the combined L-coupler technology
presented here as a viable building block for scaling modular systems.

\section*{Methods}

\subsection*{Cable connections}
The superconducting cable is a $2.05$~mm outer-diameter aluminum-aluminum coax with
low-density PTFE supplied by Hqubit System Co., Ltd.~\cite{hermercs_hqubit}. We add indium
solder to the cable in a way similar to Martin \textit{et al.}~\cite{martin2024indium}, though
we are investigating more manufacturable methods. A large modular system has to be assembled, tested and
repaired, so a connection that can be remated without added loss is what makes a failure at
module level recoverable rather than terminal. Early testing indicates this indium bond has
promise as a rematable contact, though we did not remate any connection in this work. We
believe the method is adaptable to other mating materials and to other variants of
superconducting coax without significant rediscovery.

\subsection*{L-qubit and i2m coupler}
The L-qubit is a grounded transmon whose asymmetric SQUID junctions limit its tuning range to
$\sim 800$~MHz. The i2m is an interference-based inductive coupler descended from the RF-SQUID
based design~\cite{chen2014gmon}, the coupler style used in previous cable-linked module
demonstrations. It employs an always-on negative geometric mutual inductance, $M_0$, which
shifts the off-state flux and coupling. For an appropriate choice of $M_0$ the off state, where
coupling is lowest, becomes first-order flux insensitive.

Flux through both the RF-SQUID and the i2m coupler changes not only the coupling between qubit
and cable but also the qubit frequency, which introduces an additional dephasing channel to the
L-qubit. Near the coupling off point the slope of the qubit-cable coupling versus external flux
is steep for the conventional RF-SQUID coupler and flat for the i2m
(Extended Data Fig. 1). Both the coupling and the qubit frequency are set by the same coupler
flux, so at that point the qubit frequency is flat as well, a substantial improvement in the
idling dephasing of the L-qubit relative to the original RF-SQUID design.

The i2m also reaches larger coupling to the cable with a smaller linear-to-junction inductance
ratio than the standard RF-SQUID approach, making it easier to fabricate and hysteresis free.
We measure a maximum qubit-cable coupling of $8$--$9$~MHz (Fig.~\ref{fig:Fig1}f), set by the
parameters chosen here rather than by any limit of the coupler design; the coupling can be
increased by changing the junction target of the i2m coupler. With an FSR of about $150$~MHz,
that coupling gives $g/\mathrm{FSR} = 0.05$--$0.06$.

To perform the measurements reported here we select the L-qubit operating frequency to avoid
collision with cable modes. While in some cases this permits an operating frequency at the
L-qubit natural upper sweet spot, in others the L-qubit is detuned and more sensitive to flux
noise. Every $T_2^{*}$ reported is measured at the operating frequency. Extended Data Table 1 gives the operating frequency, coherence and single-qubit gate fidelity
of every L-qubit, together with the cable standing mode each one transfers through, and
Extended Data Fig. 2 gives $T_1$ and $T_2^{e}$ for every qubit used in this work.

\subsection*{Cooldown field and the cable loss mechanism}
To characterize the field strength to which the cables are sensitive we performed cooldown
experiments using a superconducting coil in a controlled environment. The cable loops over the
coil three times, with about $5$~cm of a $1$~m cable most affected by the coil field. We apply
coil current to generate a static magnetic field while the fridge base temperature is above the
cable $T_c$ (aluminum $T_c \approx 1.2$~K). The fridge and cable are then cooled to $20$~mK
with the field maintained while data is collected. Every $Q_i$ we report for a cable or for the
standalone tester is measured at the few-photon level, low enough that the two-level-system
bath degrading $Q_i$ is left unsaturated. In a separate test we changed the applied field after
the cable had cooled below $T_c$ and found no effect on $Q_i$.

The intermediate state in type-I superconductors occurs when the applied field exceeds the
critical field for certain regions of the superconductor~\cite{schmidt}, creating a mixture of
normal and superconducting regions in which the field at the normal regions equals the critical
field. The data indicate that the cable, even at very modest fields, does not return to a full
Meissner state when cooled to $20$~mK. The geometric correction factor of $0.5$ for the
critical field of a cylinder in a transverse field~\cite{schmidt} cannot explain the impact of
such low fields, and the mechanism behind the observed loss is yet to be determined. A possible
explanation is the movement of trapped magnetic field, similar to observations in thin-film
superconducting resonators~\cite{song2009vortices,checchin2016quench}. The same dependence was
observed for aluminum coaxial cables from other vendors.

The mitigation is a magnetic-shielding braid (AA Cable Shield, Magnetic Shield
Corporation~\cite{magshield2026aacable}) around the coax. The assembled dual-Nighthawk setup is
not a like-for-like comparison with the single shielded test cable: it carries all four cables
inside the same braid, which forces the layers open, and mechanical deformation of
high-permeability shielding is known to reduce its effectiveness. Bundling is therefore a
leading candidate for the lower $Q_i$ inferred in that assembly.

\subsection*{Cable quality factor measurements}
The standalone tester places the two joins near the midpoint of the resonator, which maximizes
their contribution and gives the largest contrast between even and odd modes of the resonator
system~\cite{martin2024indium}; one end is weakly coupled to the common $Q_i$ test setup
through about $2$~fF. We see no significant
discrepancy between the two mode families. The two samples are nominally identical builds yet
differ across the band (Fig.~\ref{fig:Fig2}b).

The four L-couplers of the dual-module assembly are characterized differently, through their
own L-qubits: we swap a single excitation from the L-qubit into the mode and follow its decay.
The cable mode $T_2^{*}$ is obtained similarly, in a Ramsey-style measurement that swaps a
superposition into the mode and follows the decay of its Bloch vector.
Both types of $Q_i$ measurement are in a similar single to few photon regime, where $Q$ is not
raised by TLS saturation.

\subsection*{Relay-based state transfer}

State transfer folds L-qubit coherence, qubit-cable coupling strength, on-chip routing, the
cable attachment and the cable itself into a single number, and it is the primitive from which
distant gates and distributed Bell state preparation are derived. Network benchmarking
(NB)~\cite{helsen2023benchmarking,heya2025rb} removes state-preparation and measurement error,
so the fidelities are comparable across L-couplers. Each link is benchmarked in both
directions and we report the average of the two (Extended Data Table 2). The transfers here
run at $g/\mathrm{FSR} = 0.05$--$0.06$; the same relay scheme works to $g/\mathrm{FSR} \approx
0.15$~\cite{heya2025rb}, verified experimentally in other test samples at similar fidelities,
and reaching it requires no change to the cable, the joint or the assembly.

\subsection*{Distant CZ gate}
Randomized benchmarking of the distant CZ gate treats the whole sequence as one two-qubit
gate, so the fidelity we report is for the complete module-to-module operation.

The gate carries the overhead of the dedicated L-qubits: the excitation must be transferred
from a lattice qubit onto an L-qubit, across the cable, and back again, so each distant CZ
spends two state transfers and three local CZ gates on what is a single operation between the
lattice qubits. A large portion of the gate error is this overhead.

The L-qubits and their couplers are kept distinct from the Nighthawk square lattice for maximum
flexibility and cross-compatibility with the Nighthawk processors available on the cloud. This
division is a design choice for convenience rather than a requirement of L-coupling.
Integrating the L-coupler elements directly into the lattice qubits would remove the two state
transfers and the local CZ gates that reach them, and would therefore substantially increase
inter-module operation fidelity.

\subsection*{Error models}
Two estimates appear in Fig.~\ref{fig:Fig3}: a coherence model for state transfer and a
constituent budget for the distant CZ gate. Both are phenomenological. Neither has a free
parameter, and in both cases the gap between the estimate and the measurement is itself the
result, since it isolates the error the modelled mechanisms do not explain.

\textit{State transfer.} The model incorporates only two sources of error, the coherence of
the L-qubits and the coherence of the L-coupler cable mode. A relay transfer holds the
excitation in just these two places, so their decay and dephasing set a floor on the transfer
error that no pulse tuning can lift. The average infidelity of one element
that holds the excitation for a time $t$ is
\begin{equation}
\begin{split}
\varepsilon_{\mathrm{el}} &= \frac{1}{2}\left[1-\frac{1}{3}\left(e^{-t/T_1}+2e^{-t/T_2^{*}}\right)\right] \\
&\simeq \frac{t}{3T_1}+\frac{t}{3T_{\varphi}},
\end{split}
\label{eq:el}
\end{equation}
with $1/T_{\varphi}=1/T_2^{*}-1/(2T_1)$. The first line is exact for this depolarizing form; the
second is its expansion to first order in $t/T$, which we quote because it makes the error
additive over elements and so lets each term be attributed separately. Here $t \ll T_1,
T_2^{*}$, since the excitation occupies each element for $128$ to $156$~ns against coherence
times of tens of microseconds, and the two forms differ by less than $1.1\%$ of the element
infidelity, well inside the measurement uncertainty. We evaluate the exact form throughout. We use $T_2^{*}$ and not $T_2^{e}$
because the transfer carries no refocusing pulse. Applying Eq.~\eqref{eq:el} to the two
elements that hold the excitation gives
\begin{equation}
\begin{split}
\varepsilon = \frac{\tau_q}{3}&\left(\frac{1}{T_1^{q}}+\frac{1}{T_{\varphi}^{q}}\right) \\
+\;\frac{\tau_c}{3}&\left(\frac{1}{T_1^{c}}+\frac{1}{T_{\varphi}^{c}}\right),
\end{split}
\label{eq:coh}
\end{equation}
with $\tau_q=\tau_c=T_g/2$ for a transfer of total duration $T_g$. The cable-mode coherence
times in Extended Data Table 2 are measured through the L-qubit, since the
mode is not directly accessible.

The model takes the excitation to spend half the transfer in the qubits and half in the cable
mode, and for the whole time the excitation is in the qubits we use the shorter $T_1$ and the
shorter $T_2^{*}$ of the L-qubit pair, so the qubit terms are the worst case for each link
rather than an average over the two. It is therefore an upper bound set by coherence alone,
with no other error accounted for.

Therefore in this model coherence accounts for $77\%$ of the measured transfer error overall
and $60$ to $89\%$ on individual L-couplers. Qubit dephasing dominates on all four, at $56$ to $76\%$ of
the modelled error, and cable $T_1$ is second at $15$ to $31\%$; qubit $T_1$ and cable
dephasing together stay below $16\%$ everywhere. The cable mode carries $18$ to $37\%$ of the
modelled error and $14$ to $33\%$ of the measured error, the largest share of the latter
falling on $240$--$247$, the best link. The residual, the measured error minus
Eq.~\eqref{eq:coh}, runs from $11\%$ of the measured error on $240$--$247$ to $40\%$ on
$242$--$245$, and it does not scale with transfer duration, since the largest residual belongs
to the shortest transfer. We attribute it to pulse nonidealities across the four independently
tuned elements, two L-qubit frequencies and two coupler fluxes.

Substituting $T_1=T_2^{*}=100$\,$\mu$s for the L-qubits and leaving the cable as measured gives
the second coherence-limited upper bound in Fig.~\ref{fig:Fig3}a. At that
point the cable mode carries $68$ to $77\%$ of what remains.

For L-qubits with $T_1 = T_2^{*} = 100$\,$\mu$s and the best cable mode recorded in this
assembly, $T_1 = 38.6$\,$\mu$s, improving the relay gate width from $296$ to $126$~ns could reduce the
error to below the $10^{-3}$ threshold. These conditions are achievable when the system is
optimized for coherence, pulse-shaping, frequency allocation, and coupling.

\textit{Distant CZ gate.} The budget for $\dcz{111}{237}$ sums the separately benchmarked
errors of the gate's constituents, the two state transfers, the three local CZ gates and the
four single-qubit gates, converted to error per gate. It assumes the constituents compose independently and carries no term for
their interaction. The sum upper-bounds the gate at
$97.34 \pm 0.05\%$ and is dominated by the two state transfers at $53\%$ and the three local CZ
gates at $40\%$, accounting for $74\%$ of the measured error. We attribute a large portion of
the discrepancy to residual population left on the L-qubits over the course of the gate by
incomplete state transfer, an effect we observe.

\subsection*{Distributed entanglement}
With the distant CZ gate we established connectivity between the L-coupler and the rest of the lattice. Next we
use that gate to distribute entangled states between the two modules. All circuits in this
section were constructed and run with Qiskit.

A Hadamard followed by one distant CNOT gate, compiled from the distant CZ gate with
single-qubit gates, prepares $\Phi^{+} = (\lvert 00\rangle + \lvert 11\rangle)/\sqrt{2}$. For a
Bell pair we quote the state fidelity $F(\Phi^{+}) = \langle \Phi^{+}\rvert \rho \lvert
\Phi^{+}\rangle$, and for a GHZ state the fidelity
from the measured populations and parity amplitude. In both cases $F > 50\%$ certifies
entanglement, and for the GHZ states genuine multipartite entanglement; every state reported
here clears that bound without readout mitigation and by a wide margin with it.

Generating both Bell pairs simultaneously, the reduction falls on $\bpair{115}{233}$, whose
mitigated fidelity drops by $0.0638 \pm 0.0047$, while $\bpair{117}{231}$ is unchanged at
$0.0084 \pm 0.0058$. Both pairs remain entangled without readout mitigation.

With the reduced wiring in this setup, the two L-couplers close a ring of six lattice qubits across the two
modules, three on each. Extending the entangled state over this ring, we generate a four-qubit GHZ state
over $Q115$, $Q116$, $Q232$ and $Q233$ at $F = 77.23 \pm 0.67\%$ without readout mitigation
($91.70 \pm 0.70\%$ mitigated), and the largest state the ring supports, a six-qubit GHZ state adding
$Q117$ and $Q231$, at $F = 67.67 \pm 0.66\%$ ($90.87 \pm 0.74\%$ mitigated).
Genuine multipartite entanglement (GME) is witnessed by $F > 50\%$, the largest overlap any biseparable
state can have with a GHZ state, and both widths clear it without readout mitigation. The six-qubit state
spans the worst-readout qubit of the ring, $Q231$, which alone accounts for $0.12$ of its raw fidelity.

Taken together the results establish the full chain as a working quantum resource, running from a lattice
qubit through its local coupler, L-qubit, joint and cable, and through the same elements in the other
module.

\subsection*{Bell state fidelity}
Three circuits measure the correlators of the Bell state of the main text, with both qubits rotated
into $X$ by a Hadamard, into $Y$ by $S^{\dagger}$ then a Hadamard, and into $Z$ by no rotation.

Those three correlators fix the fidelity to all four Bell states at once, since
\begin{equation}
\begin{split}
\lvert \Phi^{\pm}\rangle &: \tfrac{1}{4}\left(II \pm XX \mp YY + ZZ\right), \\
\lvert \Psi^{\pm}\rangle &: \tfrac{1}{4}\left(II \pm XX \pm YY - ZZ\right),
\end{split}
\label{eq:bellproj}
\end{equation}
so the prepared state is identified rather than assumed. The three correlators give the fidelity as
\begin{equation}
F(\Phi^{+}) = \tfrac{1}{4}\left(1 + \langle XX\rangle - \langle YY\rangle + \langle ZZ\rangle\right) ,
\label{eq:bellfid}
\end{equation}
and $F > 50\%$ certifies entanglement, since $\tfrac{1}{2}$ is the largest overlap with
$\Phi^{+}$ any separable state can reach~\cite{guhne2009entanglement}.

\subsection*{GHZ circuit and parity oscillation}
Both states are prepared from $Q115$ as a tree rooted at one end of the bridge, with the distant gate
fired first so that the two chains fan out in parallel: three CNOT gates in two layers at four qubits, five
in three layers at six.

We measure the state by parity
oscillation~\cite{sackett2000four,monz201114qubit}, applying a collective $\pi/2$ rotation
$R(\pi/2,\varphi) = R_z(\varphi) R_x(\pi/2) R_z(-\varphi)$ to all six qubits as one physical
pulse between two virtual frame changes, and sampling 16 phases over $[0,2\pi)$. The parity
amplitude $A$ at $\nu = N$ gives the coherence $\lvert \rho_{0\ldots0,1\ldots1}\rvert = A/2$,
and with the populations from a separate circuit without the rotation it gives $F =
(P_{0\ldots0} + P_{1\ldots1} + A)/2$. The measured $\nu = 1$ harmonic is smaller than the
$\nu = N$ amplitude by a factor of $140$ at four qubits and $85$ at six, and both states satisfy
the positivity bound $\lvert \rho_{0\ldots0,1\ldots1}\rvert \le
\sqrt{P_{0\ldots0}P_{1\ldots1}}$ (Extended Data Table 3).

\subsection*{Readout mitigation}
Mitigated values use the standard correlated readout mitigation, the full $2^{N} \times 2^{N}$ assignment
matrix inverted and projected back onto the probability simplex~\cite{bravyi2021readout}, as
implemented in Qiskit Experiments. Error bars on
mitigated quantities include the shot-noise inflation of the inversion. Readout error is the
dominant limit on every entangled state reported here. This readout quality reflects the
measurement chain of this particular test system and its electronics; it is not an intrinsic
property of these qubits or the L-couplers.

\begin{acknowledgments}
We thank the broader IBM Research quantum team, with special thanks to the fabrication and integration
teams, including those former colleagues now at Anderon, and the Yorktown model shop.
\end{acknowledgments}

\bibliographystyle{apsrev4-1}
\bibliography{refs_nature}

\clearpage
\edlabels
\setcounter{topnumber}{4}
\setcounter{dbltopnumber}{4}
\setcounter{totalnumber}{6}
\renewcommand{\dbltopfraction}{0.95}
\renewcommand{\dblfloatpagefraction}{0.50}
\renewcommand{\textfraction}{0.05}

\begin{figure*}[htbp]
    \section*{Extended Data}
    \centering
    \includegraphics[width=0.62\linewidth]{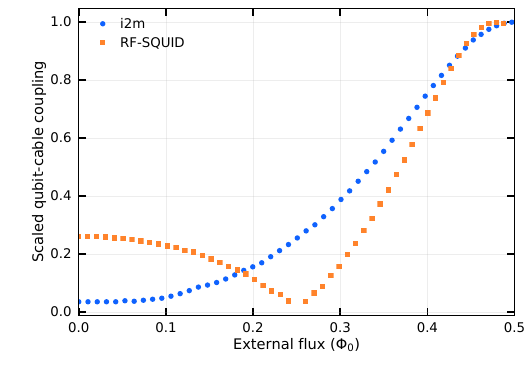}
    \caption{\textbf{Simulated i2m coupler flux response.}
Simulated qubit-cable coupling, scaled to its maximum, against external flux for the i2m
coupler (circles) and a conventional RF-SQUID coupler (squares). Near the coupling off point
the slope is steep for the RF-SQUID coupler and flat for the i2m, which is what makes the i2m
off state first-order flux insensitive. The measured coupling of one i2m coupler on this
device is given in Fig.~\ref{fig:Fig1}f.}
    \label{fig:edfig3}
\end{figure*}

\begin{figure*}[htbp]
    \centering
    \includegraphics[width=1\linewidth]{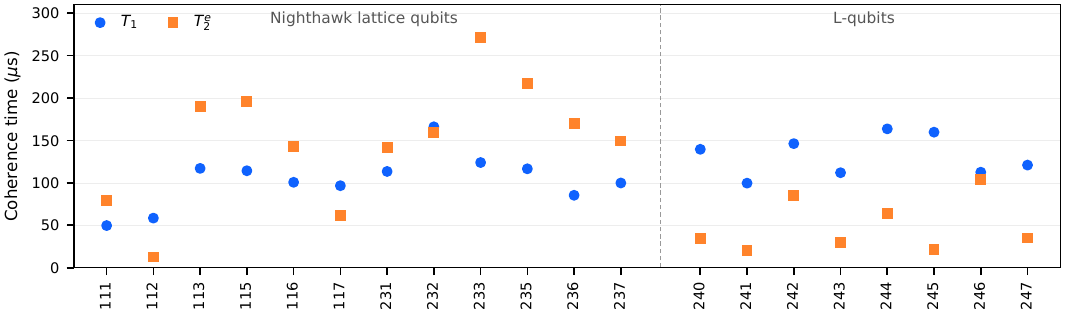}
    \caption{\textbf{Qubit coherence.} $T_1$ (circles) and
$T_2^{e}$ (squares) for every qubit used in this work, lattice qubits on the left of the
divider and L-qubits on the right.}
    \label{fig:edfig1}
\end{figure*}

\begin{table*}[htbp]
\centering
\caption{\textbf{L-qubit properties and cable modes.} L-qubit benchmarks and the cable standing mode each one transfers through. The L-qubits are
frequency-tunable; every value is quoted at the operating frequency.}
\label{tab:edqubits}
\begin{ruledtabular}
\begin{tabular}{lccccc}
L-qubit & Operating freq. (GHz) & Cable mode (GHz) & $T_1$ ($\mu$s) & $T_2^{*}$ ($\mu$s) & EPG(simul) \\
\colrule
Q240 & 5.0759 & 4.9960 & $139.6 \pm 12.2$ & $14.0 \pm 0.78$ & 0.00058 \\
Q241 & 5.0571 & 4.9770 & $99.7 \pm 5.5$ & $7.1 \pm 0.64$ & 0.00301 \\
Q242 & 5.1427 & 4.9120 & $146.3 \pm 4.5$ & $51.0 \pm 1.5$ & 0.00178 \\
Q243 & 5.1442 & 5.0880 & $112.0 \pm 3.3$ & $14.0 \pm 0.87$ & 0.00040 \\
Q244 & 5.0129 & 5.0880 & $163.7 \pm 7.9$ & $7.7 \pm 0.72$ & 0.00053 \\
Q245 & 4.9901 & 4.9120 & $159.8 \pm 6.6$ & $11.0 \pm 0.56$ & 0.00060 \\
Q246 & 5.0870 & 4.9770 & $112.6 \pm 5.4$ & $60.0 \pm 2.3$ & 0.00092 \\
Q247 & 5.0689 & 4.9960 & $121.0 \pm 4.6$ & $24.0 \pm 0.61$ & 0.00044 \\
\end{tabular}
\end{ruledtabular}
\end{table*}
\begin{table*}[htbp]
\centering
\caption{\textbf{State transfer and coherence-model inputs.} Relay-based state transfer across the four L-couplers, measured in both directions, with the
cable-mode inputs to Eq.~\eqref{eq:coh} and the coherence-limited fidelity it gives. The L-qubit inputs are
the shorter $T_1$ and the shorter $T_2^{*}$ of the pair, taken from Extended Data Table 1, which need not
belong to the same qubit.}
\label{tab:edtransfer}
\setlength{\tabcolsep}{5pt}
\begin{ruledtabular}
\begin{tabular}{lccccccc}
 & \multicolumn{3}{c}{Measured} & & \multicolumn{2}{c}{Cable mode ($\mu$s)} & \\
\cline{2-4} \cline{6-7}
L-qubit pair & L$\rightarrow$R (\%) & R$\rightarrow$L (\%) & $F$ (\%) & $T_g$ (ns) & $T_1$ & $T_2^{*}$ &
Coh.-limited $F$ (\%) \\
\colrule
Q240--Q247 & $0.635 \pm 0.030$ & $0.777 \pm 0.021$ & $99.29 \pm 0.07$ & 312 & $28.8 \pm 0.8$ & $36 \pm 3.2$ & 99.37 \\
Q241--Q246 & $1.180 \pm 0.053$ & $1.090 \pm 0.042$ & $98.87 \pm 0.05$ & 296 & $38.6 \pm 0.8$ & $51 \pm 3.0$ & 99.12 \\
Q242--Q245 & $1.070 \pm 0.030$ & $0.955 \pm 0.037$ & $98.99 \pm 0.06$ & 256 & $22.7 \pm 0.3$ & $37 \pm 1.7$ & 99.39 \\
Q243--Q244 & $1.090 \pm 0.062$ & $1.200 \pm 0.074$ & $98.86 \pm 0.06$ & 312 & $24.8 \pm 1.2$ & $37 \pm 3.5$ & 99.06 \\[4pt]
Mean & & & 99.00 & & & & \\
\end{tabular}
\end{ruledtabular}
\end{table*}
\begin{table*}[htbp]
\centering
\caption{\textbf{GHZ state measurements.} GHZ states on the ring at four and six qubits, with and without readout mitigation described
in Methods. $A$ is the parity amplitude, $\lvert
\rho_{0\ldots0,1\ldots1}\rvert = A/2$ the coherence, and the last row gives that coherence as a
percentage of the positivity ceiling $\sqrt{P_{0\ldots0}P_{1\ldots1}}$, which every reported
value satisfies. The two widths are separate measurements.}
\label{tab:edghz}
\begin{ruledtabular}
\begin{tabular}{lcccc}
 & \multicolumn{2}{c}{Four qubits} & \multicolumn{2}{c}{Six qubits} \\
\cline{2-3} \cline{4-5}
 & Unmitigated & Mitigated & Unmitigated & Mitigated \\
\colrule
$A$ ($\nu = N$) & $0.6948 \pm 0.0078$ & $0.8884 \pm 0.0081$ & $0.5690 \pm 0.0078$ & $0.8716 \pm 0.0086$ \\
$\nu = 1$ harmonic & $0.0020$ & & $0.0040$ & \\
$F$ (\%) & $77.23 \pm 0.67$ & $91.70 \pm 0.70$ & $67.67 \pm 0.66$ & $90.87 \pm 0.74$ \\[4pt]
$\lvert \rho_{0\ldots0,1\ldots1}\rvert = A/2$ & $0.3474$ & $0.4442$ & $0.2845$ & $0.4358$ \\
Coherence, \% of ceiling & & $94.3$ & & $92.5$ \\
\end{tabular}
\end{ruledtabular}
\end{table*}

\end{document}